\documentclass[aps,prd,twocolumn,showpacs,superscriptaddress,groupedaddress,nofootinbib]{revtex4}

\usepackage{amsmath}
\usepackage{amssymb}
\usepackage{amsthm}
\usepackage{bm}
\usepackage{dcolumn}
\usepackage{graphicx}
\usepackage{subfigure}

\begin{document}

\title{Binary Neutron Star Mergers: Dependence on the Nuclear Equation of State}
\author{Kenta Hotokezaka}
\affiliation{Department of Physics,~Kyoto~University,~Kyoto~606-8502,~Japan}

\author{Koutarou Kyutoku}
\affiliation{Yukawa Institute for Theoretical
Physics,~Kyoto~University,~Kyoto~606-8502,~Japan}

\author{Hirotada Okawa}
\affiliation{Yukawa Institute for Theoretical
Physics,~Kyoto~University,~Kyoto~606-8502,~Japan}

\author{Masaru Shibata}
\affiliation{Yukawa Institute for Theoretical
Physics,~Kyoto~University,~Kyoto~606-8502,~Japan}

\author{Kenta Kiuchi}
\affiliation{Yukawa Institute for Theoretical
Physics,~Kyoto~University,~Kyoto~606-8502,~Japan}

\begin{abstract}
We perform a numerical-relativity simulation for the merger of binary
neutron stars with 6 nuclear-theory-based equations of state (EOSs)
described by piecewise polytropes.  Our purpose is to explore the
dependence of the dynamical behavior of the binary neutron star merger
and resulting gravitational waveforms on the EOS of the
supernuclear-density matter.  The numerical results show that the
merger process and the first outcome are classified into three types;
(i) a black hole is promptly formed, (ii) a short-lived hypermassive
neutron star (HMNS) is formed, (iii) a long-lived HMNS is formed.  The
type of the merger depends strongly on the EOS and on the total mass
of the binaries.  For the EOS with which the maximum mass is larger
than $2M_{\odot}$, the lifetime of the HMNS is longer than $10$~ms for
a total mass $m_{0}=2.7M_{\odot}$.  A recent radio observation
suggests that the maximum mass of spherical neutron stars is
$M_{\rm{max}}\geq 1.97\pm 0.04M_{\odot}$ in one $\sigma$ level.  This
fact and our results support the possible existence of a HMNS soon
after the onset of the merger for a typical binary neutron star with
$m_{0}=2.7M_{\odot}$.  We also show that the torus mass surrounding
the remnant black hole is correlated with the type of the merger
process; the torus mass could be large, $\geq 0.1M_{\odot}$, in the
case that a long-lived HMNS is formed.  We also show that
gravitational waves carry information of the merger process, the
remnant, and the torus mass surrounding a black hole.

\end{abstract}
\maketitle

\section{Introduction}
The coalescence of binary neutron stars is one of the most promising
sources for kilometer-size laser-interferometric gravitational-wave
detectors such as LIGO~\cite{bib03}, VIRGO~\cite{bib04}, and
GEO600~\cite{bib05}.  A statistical study based on the stellar
evolution synthesis (e.g., Ref.~\cite{bib51}) suggests that detection
rate $\sim 10$ $\rm{yr}^{-1}$ may be achieved by advanced detectors
such as advanced LIGO~\cite{bib06}, advanced VIRGO~\cite{bib07}, and
LCGT~\cite{bib08}, which will be in operation in this decade.  The
merger of binary neutron stars will be a viable laboratory for
studying supernuclear-density matter through gravitational-wave
observations.  For clarifying the nature as the sources of
gravitational waves and for extracting their physical information,
theoretical templates of gravitational waves are needed.  Because the
merger is a highly dynamical process and proceeds in strong
gravitational fields, numerical relativity is the unique way for
deriving the templates of gravitational waves.

The merger of binary neutron stars also has been proposed as a likely
candidate for the central engine of short $\gamma$-ray bursts
(GRBs)~\cite{bib09, bib10}.  The observations have shown that the
central engine supplies a large amount of energy $\gtrsim 10^{48}$
ergs in a short time scale $\lesssim 2$ s~\cite{bib11}.  According to
a standard scenario based on the merger hypothesis, a stellar-mass
black hole surrounded by a hot massive torus should be formed after
the merger.  Possible relevant processes to extract the energy of this
black hole-torus system for launching a relativistic jet are
neutrino-antineutrino pair annihilation and/or magnetically driven
mechanisms.  Recent numerical studies (e.g., Ref.~\cite{bib12})
suggest that if the torus has a mass $\gtrsim 0.1M_{\odot}$, it could
supply the required energy of short GRBs by the neutrino mechanism.
The amount of the mass of the remnant torus depends on the dynamical
behavior of the merger process of binary neutron stars.  Thus, the
issue is to clarify how the formation process of a massive torus
depends on the dynamics of the merger, on the equation of state (EOS)
of neutron stars, and on parameters of the binary such as total
mass and mass ratio, for understanding the formation mechanism of
the central engine of short GRBs.

The supernuclear-density EOS plays a key role for determining the
merger process of binary neutron stars.  For example, for a given
mass, the outcome of the merger depends strongly on the EOS: For soft
EOSs, the merger results in prompt formation of a black hole.  On the
other hand, for stiff EOSs, a hypermassive neutron star (HMNS) is
formed~\cite{bib30, bib31, bib29}.  However, the actual
supernuclear-density EOS is still unknown because of our poor
knowledge about the properties of the matter above the nuclear
density.

In this paper, we extend the previous works for a long-term simulation
of binary neutron stars (e.g., Refs.~\cite{bib36,bib37,bib38}) from the
following motivation.  Recently, a piecewise-polytropic EOS for the
cold EOS, based on the nuclear theoretical calculations, was proposed
by Read \textit{et al.}, and by $\ddot{\rm{O}}$zel and
Psaltis~\cite{bib13, bib39}.  With only four parameters, the
piecewise-polytropic EOS can approximate any candidate of the EOS of
supernuclear-density matter.  By using this EOS, we can systematically
study the effects of the possible EOSs on phenomena related to the
supernuclear-density matter, such as the merger of binary neutron
stars.

In this work, we report our latest numerical-simulation results for
the merger of equal-mass binary neutron stars, for which the total
masses are in the range of $2.7M_{\odot}$--$3.0M_{\odot}$.  To
systematically study the effects of the EOS on the merger, we use 6
different stiff EOSs which are described in Ref.~\cite{bib13}.  In the
present simulation, we follow the inspiral motion for 5--7 orbits and
the merger process up to formation of a stationary black hole or HMNS.
After a black hole is formed, we study the dependence of the
properties of the resulting torus, such as its mass, on the EOS and
the total mass of the binaries.  We also extract gravitational waves
and show the dependence of the gravitational waveforms and their
spectra on the EOS.

The paper is organized as follows. In Sec.~II, we summarize the
formulation and numerical schemes employed in our numerical code {\tt
SACRA}, and the EOS employed in this study.  In Sec.~III.A, we
describe our simulation results for the typical cases of the merger.
We define three types of the merger process, which clearly classify
the dependence of the dynamical behavior of the merger on the EOSs.
In Sec.~III.B, we summarize the characteristic features of
gravitational waveforms for each type.  Section~IV is devoted to a
summary.  Throughout this paper we use the geometrical units of
$c=G=1$ where $c$ and $G$ are the speed of light and gravitational
constant, respectively.

\section{Formulation}

\subsection{Numerical method}
We follow the late inspiral and merger phases of binary neutron stars
using a numerical-relativity code, called {\tt SACRA}, described in
Ref.~\cite{bib01}.  {\tt SACRA} employs a moving puncture version of
the Baumgarte-Shapiro-Shibata-Nakamura formalism~\cite{bib32,
bib33,bib15} to solve Einstein's evolution equation without imposing
any symmetry.  In {\tt SACRA}, we evolve a conformal factor $W\equiv
\gamma^{-1/6}$, the conformal three metric
$\tilde{\gamma}_{ij}=\gamma^{-1/3}\gamma_{ij}$, the trace of extrinsic
curvature $K$, the conformal trace-free extrinsic curvature
$\tilde{A}_{ij}=\gamma^{-1/3}\left( K_{ij} - K\gamma_{ij}/3
\right)$, and an auxiliary variable $\tilde{\Gamma}^{i}\equiv
-\partial_{j}\tilde{\gamma}^{ij}$.  Here $\gamma_{ij}$ is the three
metric, $K_{ij}$ is the extrinsic curvature, and $\gamma \equiv
\rm{det}(\gamma_{ij})$.  In the numerical simulation, a fourth-order
finite differencing scheme in space and time is used implementing an
adaptive mesh refinement (AMR) algorithm (at refinement boundaries, a
second-order interpolation scheme is partly adopted).  The advection
terms such as $\beta^{i}\partial_{i}\tilde{\gamma}_{jk}$ are evaluated
by a fourth-order non-centered finite difference~\cite{bib16}.  A
fourth-order Runge-Kutta method is employed for the time evolution.

Following Ref.~\cite{bib16}, we adopt a moving-puncture gauge
condition as
\begin{eqnarray}
&&(\partial_{t}-\beta^{j}\partial_{j})\beta^{j}=0.75B^{j}, \\
&&(\partial_{t}-\beta^{j}\partial_{j})B^{i}
=(\partial_{t}-\beta^{j}\partial_{j})\tilde{\Gamma^{i}}-\eta_{s}B^{i},
\end{eqnarray}
where $B^{i}$ is an auxiliary variable and $\eta_{s}$ is an arbitrary
constant.  In the present paper, we set $\eta_{s} \simeq 3/M$. Here,
\textit{M} denotes the mass for each neutron star in isolation.

For the hydrodynamics, we employ a high-resolution central scheme by
Kurganov and Tadmor~\cite{bib35} with a third-order piecewise
parabolic interpolation and with a steep min-mod limiter.

In {\tt SACRA}, an AMR algorithm is adopted (see Ref.~\cite{bib01} for
details).  In the present work, we prepare seven refinement levels
both to accurately resolve the structure of two neutron stars and to
extract gravitational waves in a local wave zone.  In our simulations,
two sets of four finer domains comoving with the neutron stars cover
the region in their vicinity.  The other three coarser domains cover
both neutron stars by a wider domain with their origins being fixed
approximately at the center of mass of the binary.  Each refinement
domain consists of the uniform, vertex-centered Cartesian grids with
($2N+1, 2N+1, N+1$) grid points for ($x, y, z$) with the equatorial
plane symmetry at $z=0$ imposed.  The half of the edge length of the
largest domain (i.e., the distance from the origin to outer boundaries
along each axis) is denoted by $L$ which is chosen to be $\agt
\lambda_{0}$, where $\lambda_{0} =\pi/\Omega_{0}$ is the initial
wavelength of gravitational waves.  The grid spacing for each domain
is then $h_{l}=L/(2^{l}N)$, where $l=0$--$6$.  In this work, we
typically choose $N=60$.  With this grid resolution, the semi-major
diameter of each neutron star is covered by about 100 grid points
(cf. Table~\ref{t5.1}).  In addition, we performed numerical simulations
with lower grid resolutions, $N=36$, 42, and 50, to check the
convergence of the numerical results.  The property of the convergence
is essentially the same as in Refs.~\cite{bib01, bib15}. 

\subsection{Models of Equation of State}

\begin{table*}
\caption[table]{Parameters of the piecewise-polytropic EOS, 
the maximum mass of spherical neutron stars, $M_{\rm{max}}$, and the
radius of a spherical neutron star of mass $M=1.4M_{\odot}$, 
$R_{1.4}$, for each EOS.  Composition means strongly interacting
components (n=neutron, p=proton, H=hyperon, Q=quark, $\pi^{0}$=pion)
and APR4, SLy, H3, H4, and ALF2 include leptonic
components.}\label{table2}
\begin{center}
 \begin{tabular}{|c|cccccc|cc|} \hline \hline
 EOS & $\log P_1 (\rm{dyne/cm}^{2}$) & $\Gamma_{1}$ & $\Gamma_{2}$ & $\Gamma_{3}$ &
$M_{\rm{max}} (M_{\odot})$ & $R_{1.4}(\rm{km})$& Approach & composition \\ 
 \hline
APR4 & 34.269 & 2.830 & 3.445 & 3.348 & 2.213 & 11.428 & Variational-method & np \\ 
SLy &34.348 & 3.005 & 2.988 & 2.851 & 2.049 & 11.736 &  Effective-one-body potential
& np \\ 
H3 & 34.646 & 2.787 & 1.951 & 1.901 & 1.788 & 13.840 & Relativistic mean field & npH
 \\ 
H4  &34.669 & 2.909 & 2.246 & 2.144 & 2.032 & 13.759 & Relativistic mean field & npH
\\ 
ALF2 &34.055 & 4.070 & 2.411 & 1.890 & 2.086 & 13.188 & APR+Quark matter & npQ\\  
PS  &34.671 & 2.216 & 1.640 & 2.365 & 1.755 & 15.472 & Pion condensation &
n$\pi^{0}$\\ \hline \hline
\end{tabular}
\end{center}
\end{table*}

The parameterized piecewise-polytropic EOS~\cite{bib13,bib39} is
useful to systematically study the dependence of the dynamical
behavior of the merger on the EOS of the supernuclear-density matter.
In this work, we employ a parameterized piecewise-polytropic EOS
proposed by Read \textit{et al.}~\cite{bib13}.  This EOS is written in
terms of four segments of polytropes
\begin{align}
P = & K_{i}\rho^{\Gamma_{i}} \\ \nonumber 
&\text{( for $\rho_{i}\leq \rho<\rho_{i+1}$, $0\leq i \leq 3$)},
\end{align}
where $\rho$ is the rest-mass density, $P$ is the pressure, $K_{i}$ is
the polytropic constant, and $\Gamma_{i}$ is the adiabatic index.  We
refer to the pressure in the form of Eq.~(3) as the cold-part
pressure, $P_{\rm{cold}}$.  At each boundary of the piecewise
polytropes, $\rho=\rho_{i}$, the pressure is required to be
continuous, i.e.,
$K_{i}\rho_{i}^{\Gamma_{i}}=K_{i+1}\rho_{i}^{\Gamma_{i+1}}$.  Read
\textit{et al.} determine these parameters in the following
manner~\cite{bib13}.  First, they fix the EOS of the crust as
$\Gamma_{0}=1.357$ and $K_{0}=3.594\times 10^{13}$ in the cgs unit.
Then they determine $\rho_{2}=1.85\rho_{\rm{nucl}}$ and
$\rho_{3}=3.70\rho_{\rm{nucl}}$ where $\rho_{\rm{nucl}}=2.7\times
10^{14}$ $\rm{g/cm^{3}}$ is the nuclear saturation density.  With this
preparation, they choose the following four parameters as a set of
free parameters: $\{P_{1}, \Gamma_{1}, \Gamma_{2}, \Gamma_{3}\}$.
Here $P_{1}$ is the pressure at $\rho=\rho_{2}$, and from this,
$K_{1}$ and $K_{i}$ are determined by
$K_{1}=P_{1}/\rho_{2}^{\Gamma_{1}}$ and
$K_{i+1}=K_{i}\rho_{i}^{\Gamma_{i}-\Gamma_{i+1}}$.  Therefore the EOS
is specified by choosing the four parameters $\{P_{1}, \Gamma_{1},
\Gamma_{2}, \Gamma_{3}\}$.

In this paper, we adopt 6 models of piecewise-polytropic EOS which
describe the following EOSs based on nuclear theoretical calculations.
\begin{enumerate}
\item APR4: derived by a variational-method with the AV18 2-body 
potential, the UIX 3-body potential, and relativistic boost
corrections (see Ref.~\cite{bib17});
\item SLy: derived by using an effective potential approach of the 
Skyrme type (see Ref.~\cite{bib18});
\item H3: derived by a relativistic mean-field approach including 
hyperons. The incompressibility, the effective mass, and the
nucleon-meson coupling are chosen to be $K=300$ MeV,
$m_{*}/m_{n}=0.7$, and $x_{\sigma}=0.6$.  Here $m_{n}$ is the
nucleon-mass (see Refs.~\cite{bib19,bib20});
\item H4: the same as H3 but for $x_{\sigma}=0.72$ 
(see Refs.~\cite{bib19,bib20});
\item ALF2: a hybrid EOS which describes nuclear matter for a low 
density and color-flavor-locked quark matter for a high density.  The
transition density and the interaction parameter are chosen to be
$\rho_{c}=3\rho_{\rm{nucl}}$ and $c=0.3$ (see Ref.~\cite{bib21});
\item PS: derived by using a potential approach. This EOS describes 
a neutron matter with pion condensation (see Ref.~\cite{bib22}).
\end{enumerate}
Table~\ref{table2} lists the parameters of piecewise-polytropic EOSs
employed in this work.  These EOSs are relatively stiff, and hence,
the maximum mass of spherical neutron stars is larger than
$1.75M_{\odot}$.  The choice of these EOSs is motivated by the recent
discovery of a heavy neutron star with mass $1.97 \pm 0.04M_{\odot}$
(one $\sigma$ error)~\cite{bib34}.  This value has become the new
standard for the minimum value of the neutron star maximum
mass~\cite{bib40, bib48}.

The thermal pressure should be taken into account for numerical
simulations, because matter in the merged neutron stars becomes hot
with temperature to $T \geq 10$ MeV due to the shock heating at the
merger (e.g., Refs.~\cite{bib02,skks11}).  In this case, the thermal
energy is not negligible.  To approximately include the thermal
pressure, we employ the EOS which is described by
\begin{eqnarray}
P(\rho, \varepsilon) = P_{\rm{cold}}(\rho) + P_{\rm{th}}(\rho, \varepsilon),
\end{eqnarray}
where $\varepsilon$ is the specific internal energy, $P_{\rm{cold}}$
is the pressure determined by the piecewise-polytropic EOS, and
$P_{\rm{th}}$ is the thermal part of the pressure which is given by
\begin{eqnarray}
P_{\rm{th}} = \left( \Gamma_{\rm{th}} - 1 \right)
 (\varepsilon-\varepsilon_{\rm{cold}}) \rho.
\end{eqnarray}
Here $\varepsilon_{\rm{cold}}$ is determined from $P_{\rm{cold}}$ by
the first law of thermodynamics~\cite{bib13}.  In our simulations, we
focus only on the case that the shock heating efficiency is relatively
low, i.e., $\Gamma_{\rm{th}}=1.357$.  To study the effect of thermal 
pressure on the outcome of the merger, we also employed 
$\Gamma_{\rm{th}} = 1.5$, $1.7$, and $1.8$ for a few simulations with 
lower grid resolutions.  We found that the outcome formed soon after 
the merger (a black hole or HMNS) depends very weakly on the value of 
$\Gamma_{\rm{th}}$ (see also Ref.~\cite{bib31}), although the
long-term evolution process of a HMNS depends on it~\cite{bib02}. 

\begin{table*}[ht]
\caption[table]{Key parameters for the initial models adopted in the 
numerical simulation. $m_{0}$ is the sum of the ADM masses of two
neutron stars in isolation ($2M$); $M_{0}^{\rm{ADM}}$ and
$J^{\rm{ADM}}_{0}$ are the ADM mass and angular momentum of the
system, respectively; $M_{*}$ is the baryon rest mass; $\Omega_{0}$ is
the angular velocity.  We also show the setup of the grid structure
for the computation with our AMR algorithm.  $\Delta x =h_{6}
=L/(2^{6}N)$ is the grid spacing at the finest resolution domain with
$L$ being the location of the outer boundaries for each axis.
$R_{\rm{diam}}$ denotes the number of the grid points assigned inside
the semimajor diameter of the neutron stars.  $\lambda_{0}$ is the
wevelength of gravitational waves of the initial configuration.  In
the last two columns, we show the simulation results for the rest mass
of the torus $M_{\rm{torus}}$ and the type of the merger process for
each model.  The rest mass of the torus surrounding the black hole is
determined at 1~ms after the black hole formation.  For APR4-27, a
HMNS with the lifetime $\gg 10$ ms is formed.  For H4-30, the type of
the merger process is ambiguous because a black hole is formed at only
1.5 ms after the onset of the merger.  }\label{t5.1}
\begin{center}
 \begin{tabular}{lcccccccccc} \hline \hline
 \textrm{Model} & $m_{0}$& $M^{\rm{ADM}}_{0}$ & $J^{\rm{ADM}}_{0}$ & $M_{*}$ &
$m_{0}\Omega_{0}$ & $\Delta x/m_{0}$ & $R_{\rm{diam}}/\Delta x$ & $L/\lambda_{0}$ &
$M_{\rm{torus}}/M_{\odot}$ & type\\ \hline
APR4-27~~ & ~2.7~ & ~2.67~ & ~7.16~ & ~3.00~ & ~0.0221~ & 0.043 & 99 & 1.16 & - &
III\\ 
APR4-28 & 2.8 &2.77 & 7.70 &3.12 & 0.0221& 0.041 &102 & 1.11 & 0.003 & I\\ 
APR4-29 & 2.9 & 2.87 & 8.26 & 3.26 & 0.0221& 0.039 & 102 & 1.05 & $<$0.001 & I\\
\hline 
SLy-27 & 2.7 & 2.67 & 7.16 &2.98 & 0.0221& 0.045 & 101 &1.21  & 0.02 & II\\
SLy-28 & 2.8 & 2.77 & 7.70 & 3.12 & 0.0221& 0.043 & 102 & 1.15 & $<0.001$ & I\\ \hline
H3-27 & 2.7 &2.68 & 7.39  & 2.94 & 0.0221 &  0.056  & 102 & 1.50 & 0.05 & II  \\
H3-29 & 2.9 & 2.87 & 8.27 & 3.18& 0.0221 & 0.050 & 103 & 1.34& 0.01 & I\\ \hline
H4-27 & 2.7 & 2.68 & 7.39 &2.94 &0.0221 & 0.056 & 103 & 1.50& 0.18 & III\\
H4-29 & 2.9 & 2.87 & 8.27 & 3.18&0.0221& 0.051 & 101 & 1.37 & 0.02 & II\\
H4-30 & 3.0 & 2.97 & 8.85 & 3.30&0.025 & 0.048 & 102 & 1.49 & 0.01 & I or II\\ \hline
ALF2-27 & 2.7 & 2.67 & 7.17 &2.98 & 0.0221 & 0.049 & 102 &1.32 & 0.16 & III\\
ALF2-29 & 2.9 & 2.87 & 8.51 &3.22 & 0.0221 & 0.045 & 102 &1.22& 0.02 & II\\
ALF2-30 & 3.0 & 2.97 & 8.85 &3.34  &0.0221 & 0.043 & 102 &1.32& 0.003 & I \\ \hline
PS-27 & 2.7 & 2.68 & 7.57 & 2.88 &0.020 & 0.073 & 92 &1.60& 0.04 & III\\ 
PS-29 & 2.9 & 2.88 & 8.73 & 3.12 & 0.020 & 0.065 & 92 & 1.48& 0.02 & II\\
PS-30 & 3.0 & 2.97 & 8.85 & 3.24 &0.025 & 0.056 & 102& 1.71& 0.01 & I\\ \hline \hline
\end{tabular}
\end{center}
\end{table*}

\subsection{Initial data}

We prepare binary neutron stars in quasiequilibrium states for the
initial condition of numerical simulations. To track more than 5
quasicircular orbits with small eccentricity for deriving accurate
gravitational waveforms in the late inspiral and merger phases, 
orbital separation of the initial configuration is chosen to be large
enough that the time scale of gravitational radiation reaction is much
longer than the orbital period of the binary. 

The formulation and methods for a solution of Einstein's constraint
equation and equations of hydrostatics are the same as those adopted
in our previous works~\cite{bib01,bib24,bib37}, except for the choice
of EOSs. We assume the conformal flatness of the three metric
$\gamma_{ij} = \psi^4 f_{ij}$, the maximal slicing condition $K=0$,
and their preservation in time. Here, $\psi$ is a conformal factor and
$f_{ij}$ is the flat spatial metric. The piecewise-polytropic EOS
described in the previous subsection is adopted to model the neutron
star matter because the fluid inside the neutron stars in the late
inspiral phase are believed to be well approximated by a cold,
zero-temperature matter. The neutron stars are assumed to have an
irrotational velocity field, which is believed to be an
astrophysically realistic configuration~\cite{bibi4,bibi5}. Numerical
computations are performed using the spectral-method library,
LORENE~\cite{bibi3}. The details of the numerical methods and the
analysis of quasiequilibrium states are summarized in
Ref.~\cite{bibi1} (see also Ref.~\cite{bibi2}). We note that the
virial error of the quasiequilibrium, which we define as the relative
difference between the ADM and Komar masses, is always smaller than
$10^{-4}$ for our numerical solutions.

\subsection{Gravitational wave extraction and waveforms}

Gravitational waves are extracted by calculating the complex Weyl
scalar $\Psi_{4}$, using the same procedure as in Ref.~\cite{bib01}.
Gravitational waveforms are calculated by
\begin{eqnarray}
h_{+}(t)-ih_{\times}(t) = -\lim_{r\rightarrow \infty}
\int^{t}dt^{\prime}\int^{t^{\prime}}dt^{\prime \prime}
\Psi_{4}(t^{\prime \prime},r).
\end{eqnarray}
Here we omit arguments $\theta$ and $\phi$.  We evaluate $\Psi_{4}$ at
a finite coordinate radius $r=400M_{\odot}\simeq 590$ km.  In the
standard spherical coordinate $(r, \theta, \phi)$, $\Psi_{4}$ can be
expanded in the form
\begin{eqnarray}
\Psi_{4}(t, r, \theta, \phi) = \sum_{lm}\Psi_{4}^{lm}(t,r)_{-2}Y_{lm}(\theta, \phi),
\end{eqnarray}
where $_{-2}Y_{lm}$ are spin-weighted spherical harmonics of weight
$-2$ and $\Psi_{4}^{lm}$ are expansion coefficients defined by this
equation.  In this work, we focus only on the $(l,|m|)=(2,2)$ mode.

We evaluate the amplitude of the Fourier spectrum of gravitational waves,
\begin{eqnarray}
\tilde{h}(f) = \sqrt{\frac{| \tilde{h}_{+}(f)| ^{2}+| \tilde{h}_{\times}(f)|^{2}}{2}},
\end{eqnarray}
where $f$ is the frequency, and $\tilde{h}_{+}(f)$ and
$\tilde{h}_{\times}(f)$ are the Fourier transformation of the plus and
cross modes of gravitational waves observed along the $z$ axis.  The
effective amplitude of gravitational waves for a given frequency is
defined by,
\begin{eqnarray}
        h_{\rm{eff}}(f) = f\tilde{h}(f).
\end{eqnarray}
Note that this is the most optimistic value for the effective
amplitude.  Actually, the amplitude of gravitational waves depends on
an angle locating the source in the sky and on an angle specifying the
orientation of orbital plane of the binary neutron star.  The angular
average of the effective amplitude is approximately $\simeq 0.4
h_{\rm{eff}}$.

\subsection{Mass, linear momenta and angular momenta}

We monitor the ADM mass $M_{\rm{ADM}}$, the linear momentum $P_{i}$,
and the angular momenta $J_{i}$ during the evolution.  These
parameters are defined by the integrals on two surfaces of a
coordinate radius $r=400$, $300$, and $240M_{\odot}$, 
\begin{eqnarray}
M_{\rm{ADM}}(r) & = & \frac{1}{16\pi}\int_{r} 
\sqrt{\gamma}\gamma^{ij}\gamma^{kl}(\gamma_{ik,j}-\gamma_{ij,k})dS_{l},\\
P_{i}(r) & =
&\frac{1}{8\pi}\int_{r}\sqrt{\gamma}(K_{i}^{j}-K\gamma^{j}_{i})dS_{j},
\\
J_{i}(r) & = &
\frac{1}{8\pi}\epsilon_{ikl}\int_{r}\sqrt{\gamma}x^{l}(K^{jk}-K\gamma^{jk})
dS_{j}, 
\end{eqnarray}
where $dS_{l}$ is the surface element and $\epsilon_{ijk}$ is the
Levi-Civita symbol.  Then, we extrapolate these quantities for $r
\rightarrow \infty$ to obtain the asymptotic value.

We also monitor the total baryon rest mass
\begin{eqnarray}
M_{*}=\int \rho u^{t}\sqrt{-g}d^{3}x,
\end{eqnarray}
where $u^{t}$ is the time-component of the four velocity, and $g$ is
the determinant of the space-time metric.  After the black hole
formation, we calculate the torus mass defined by
\begin{eqnarray}
M_{\rm{torus}}=\int_{r>r_{\rm{AH}}} \rho u^{t}\sqrt{-g}d^{3}x,
\end{eqnarray}
where $r_{\rm{AH}}$ is the coordinate radius of the apparent horizon.

\begin{figure*}[ht]
\begin{center}
\begin{tabular}{l l}
\rotatebox{0}{\includegraphics[scale=0.9,clip]{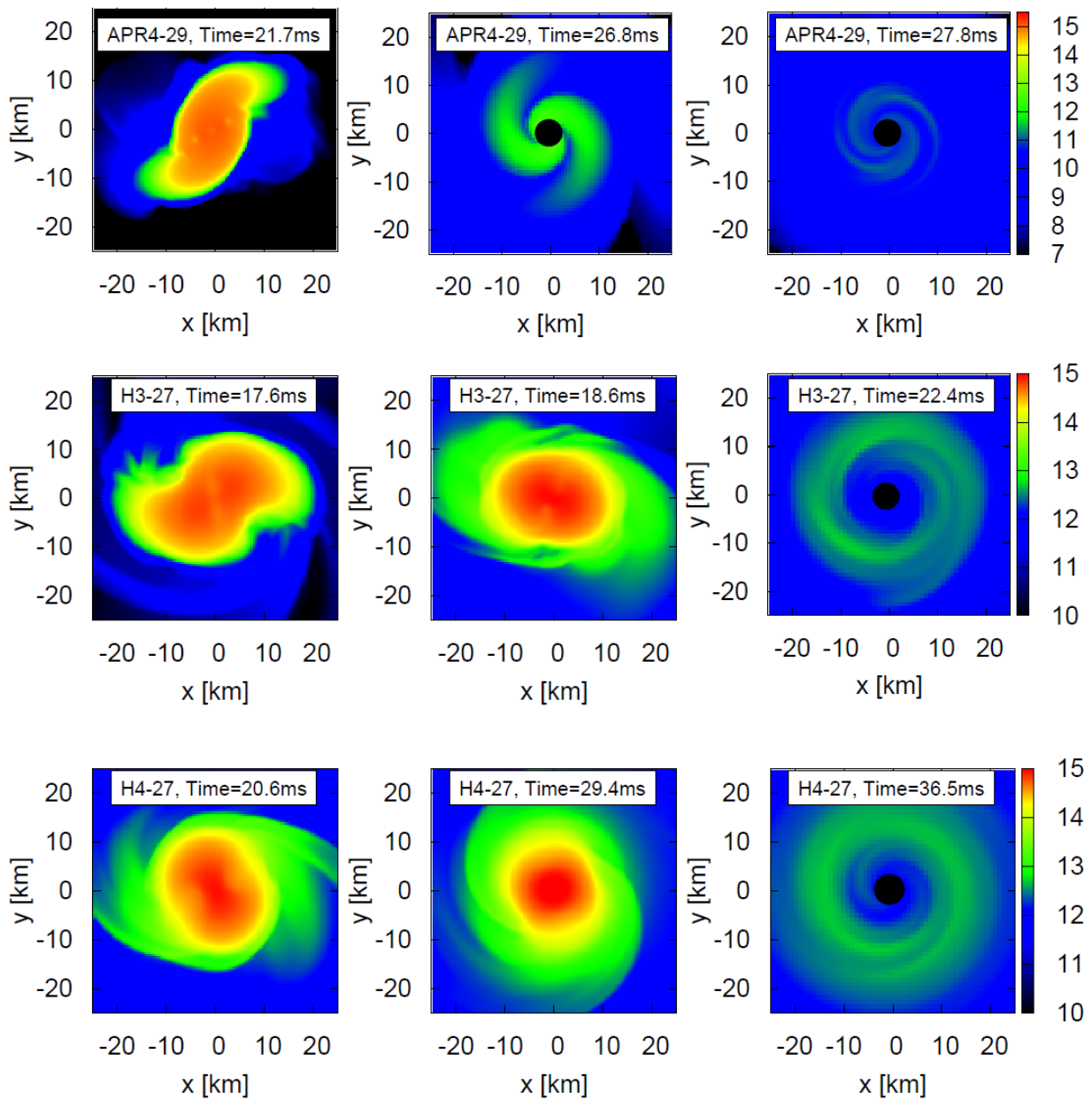}}\\
\end{tabular}
\caption{Colormap of the density, $\log \rho$ $(\rm{g/{cm}^{3}})$.
\textit{Top}, \textit{middle}, and \textit{bottom} rows show the 
snapshots for APR4-29, H3-27, and H4-27, respectively.  The black
filled circle denotes the region inside apparent horizon.  Note that
the density range of the color bar for APR4-29 is different from the
other models.}
\label{snapshot}
\end{center}
\end{figure*}

\section{Numerical Results}

Table~\ref{t5.1} lists the numerical models adopted in this paper.
The simulations were performed from 5 -- 7 orbits before the onset of
the merger to $3$~ms after the formation of a black hole or to the phase
in which a HMNS relaxes to a quasi-stationary state.

\subsection{Dynamics}

The merger dynamics are determined primarily by three forces: gravity,
pressure, and centrifugal forces.  If its gravity is stronger than the
other forces, the merged neutron star collapses to a black hole soon
after the onset of the merger.  On the other hand, if the other forces
could overcome the gravity in a compact state, the core of the
merged neutron star bounces and a rapidly rotating and oscillating
HMNS is formed.  This dynamical behavior depends on the EOS and on the
total mass of the binary neutron star, $m_{0}$.  In the following, we
show the results of our numerical simulations focusing on the
dynamical behavior of the merger.

\subsubsection{\textit{Classification of the merger process}}

It is natural to expect that a black hole is eventually formed after a
merger of any binary neutron star, because a typical total mass of
binary neutron stars is $\sim 2.7M_{\odot}$~\cite{bib41}, which is
likely to exceed the maximum mass of spherical neutron stars.
Although binary neutron stars have such a large total mass, a HMNS is
often formed as a transient outcome of the merger, which is supported
by the strong centrifugal force caused by rapid and differential
rotation as well as by thermal pressure.  Thus, one can classify the
merger process and the resulting remnant of binary neutron stars into
three types (see Table II):
\begin{itemize}
\item type I: A black hole is promptly formed;
\item type II: A short-lived HMNS is formed \\
\hspace{4.5cm}$(\tau_{H}<5$ $\rm{ms})$; 
\item type III: A long-lived HMNS is formed\\
\hspace{4.5cm} $(\tau_{H}>5$ $\rm{ms})$.
\end{itemize}
Here $\tau_{H}$ is a lifetime of a HMNS.  We note that a massive and
stable rigidly rotating neutron star may be formed if the total mass
is only slightly larger than the maximum mass of spherical neutron
stars or the contribution of the thermal pressure is significant.
However, we do not consider this fourth possibility in this paper (but
see Ref.~\cite{skks11}).

Figure~\ref{snapshot} shows the snapshots of the density colormap for
three types of the merger process.  Since the dynamics of the binary
neutron stars in the inspiral phase is similar among three types, we
focus only on the dynamics after the onset of the merger in the
following. 

\textit{Top} panels; APR4-29 (type I).
Soon after the onset of the merger (Fig.~\ref{snapshot}, \textit{top
left}), the merged object collapses promptly to a black hole and tiny
materials remain outside the black hole (Fig.~\ref{snapshot},
\textit{top center}).  The black filled circle denotes the inner
region of the apparent horizon.  Note that the spiral arms are formed
and the materials in their outer region obtain angular momentum from
the materials in the inner region by gravitational torques resulting 
from the non-axisymmetric structure.  At $\simeq 5$ ms after the onset
of the merger, a quasi-stationary torus is formed with the maximum
density $\rho_{\rm{max}} \sim 10^{11.5}$ $\rm{g/cm^{3}}$, and spreads
to about $10$ km from its center (Fig.~\ref{snapshot}, \textit{top
right}).  However the torus mass is small $\lesssim 10^{-3}M_{\odot}$.

\textit{Middle} panels; H3-27 (type II). After the onset of the 
merger, the merged core bounces due to strong centrifugal and pressure
forces.  The resulting HMNS has a double-core structure in which they
rotate around each other (Fig.~\ref{snapshot}, \textit{middle
center}).  At $\simeq 5$ ms after the onset of the merger, the HMNS
collapses to a black hole (Fig.~\ref{snapshot} \textit{middle right}),
because its angular momentum decreases due to the emission of
gravitational waves.  We find that the collapse of the HMNS occurs
during the phase in which the HMNS has a non-axisymmetric shape.  The
resulting torus around the black hole (Fig.~\ref{snapshot}
\textit{middle right}) spreads to about $20$ km.  This torus has the
maximum density $\rho_{\rm{max}} \sim 10^{12.5}$ $\rm{g/cm^{3}}$.  In
this case, the torus mass is $\simeq 0.05M_{^\odot}$.  The resulting
black hole-torus system is a candidate for the central engine of short
GRBs.  

For H4-29, ALF2-29, and PS-29, a HMNS of double-core structure is also
formed while for SLy-27 a HMNS of ellipsoidal shape is formed.
Irrespective of the configuration of the HMNS, however, the mass of
remnant torus formed after the black hole formation is $\simeq
0.02M_{\odot}$ and much less than that for H3-27.  This suggests that
the torus mass may be suppressed for a massive system, for which the
total mass is close to $M_{\rm{crit}}$; see Eq.~(15) for the
definition of $M_{\rm{crit}}$.

\begin{figure}[t]
 \begin{center} \includegraphics[scale =0.7]{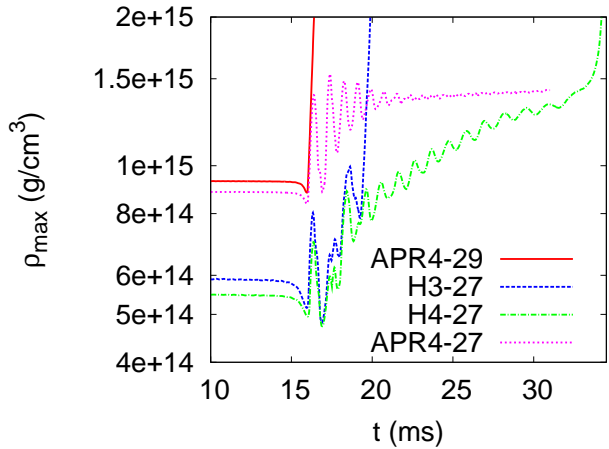} \end{center}
 \caption{The evolution of the maximum baryon rest-mass density,
 $\rho_{\rm{max}}$, for three models.  The solid, dashed, dash-dotted,
 and dotted curves denote the results for models APR4-29 (type I),
 H3-27 (type II), H4-27 (type III), and APR4-27 (type III),
 respectively.  } \label{rho}
\end{figure}

\begin{figure*}[htbp]
 \begin{center}
 \includegraphics[scale =0.6,clip]{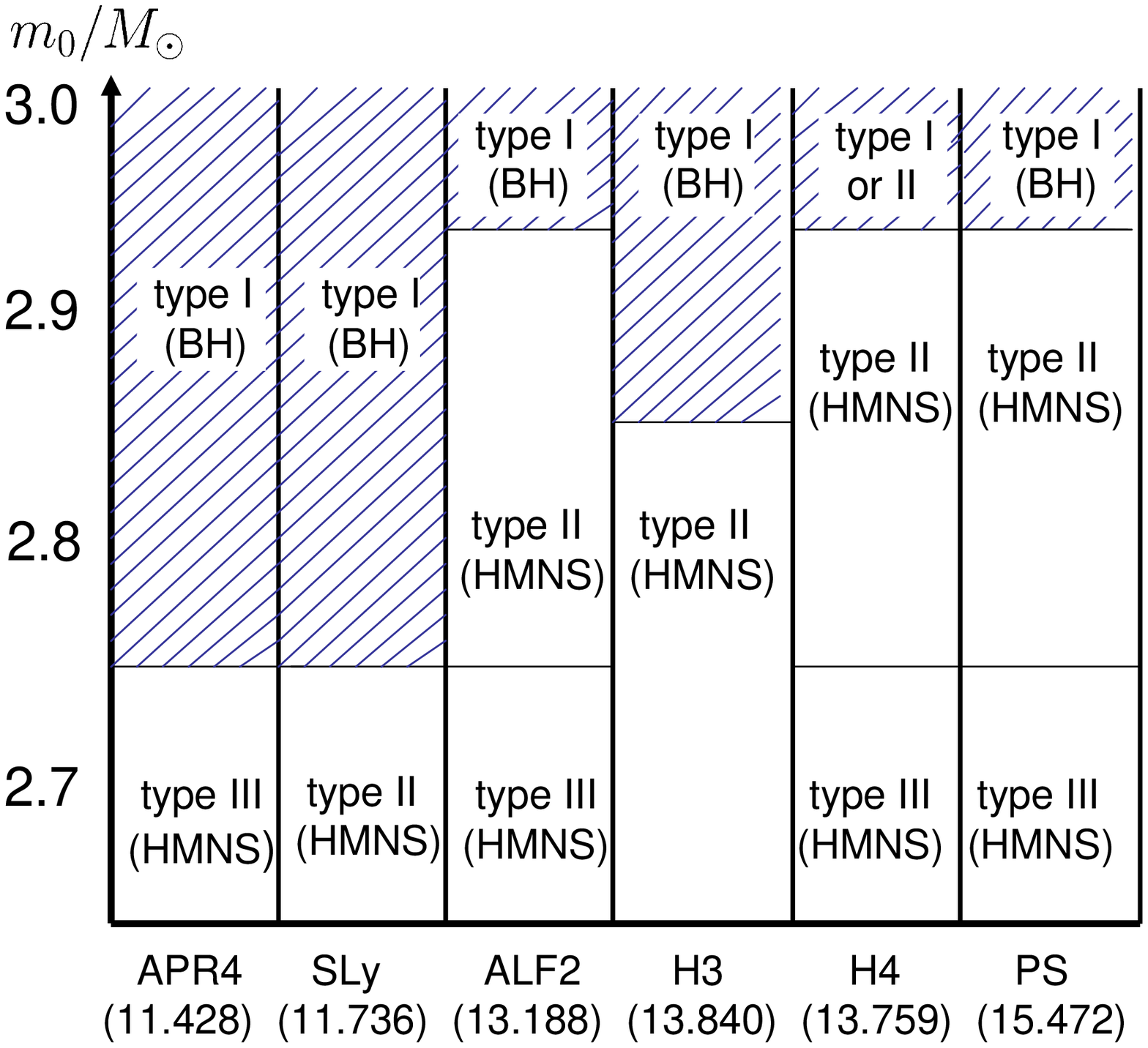}
 \end{center}
\caption{Type of the merger process and the remnants for each model. 
The vertical axis shows the total mass of two neutron stars.  The
 horizontal axis shows the EOSs together with their radii for
 $M=1.4M_{\odot}$, $R_{1.4}$~ km.  }\label{f5.8} \label{type}
\end{figure*}

\textit{Bottom} panels; H4-27 (type III). 
After the onset of the merger, a HMNS, which has a double-core
structure, is formed as in the case of H3-27 (Fig.~\ref{snapshot},
\textit{bottom left}).  Because the lifetime of the HMNS is
sufficiently long, a large amount of angular momentum is dissipated by
gravitational waves.  As a result, the HMNS approaches an axisymmetric
ellipsoidal shape (Fig.~\ref{snapshot}, \textit{bottom center}).
After the significant loss of the angular momentum, it collapses
eventually to a black hole (Fig.~\ref{snapshot}, \textit{bottom
right}).  The resulting torus surrounding the formed black hole has
the maximum density $\rho_{\rm{max}} \sim 10^{13}$ $\rm{g/cm^{3}}$ and
spreads widely to $r\sim 25$ km.  In this case, the torus mass is
$\simeq 0.18M_{\odot}$.  The resulting black hole-torus system may be
regarded as a promising candidate for the central engine of short 
GRBs.  

APR4-27, ALF2-27, and PS-27 also show the merger process of type III.
For ALF2-27, the mass of the remnant torus is larger than
$0.1M_{\odot}$ as for H4-27.  This suggests that with the stiff EOSs
for which the maximum mass of spherical neutron stars is larger than
$2M_{\odot}$, a massive torus could be the outcome when the total mass
of the binary is $\approx 2.7M_{\odot}$.  For PS-27 with which the
maximum mass of spherical neutron stars is $1.75M_{\odot}$, despite
type III, the mass of the remnant torus is much less ($\simeq
0.04M_{\odot}$).  This indicates that for the EOS with small maximum
mass, the torus mass may be suppressed.  More detailed reasons that
could cause the difference in the remnant torus mass will be discussed
in Sec. III.A.3.  For APR4-27, a quasi-stationary HMNS is formed.
Because the degree of differential rotation is still high and the mass
is much higher than the maximum mass of spherical neutron stars, it
will eventually collapse in the presence of the magnetic field or
viscosity in reality (see Ref.~\cite{bib47}).

Figure~\ref{rho} plots the evolution of the maximum baryon rest-mass
density, $\rho_{\rm{max}}$, for models (types) APR4-29 (type I), H3-27
(type II), H4-27 (type III), and APR4-27 (type III).  For APR4-29, for
which a black hole is formed promptly, $\rho_{\rm{max}}$ increases
monotonically after the onset of the merger.  For H3-27 and H4-27 for
which a HMNS is formed transiently, $\rho_{\rm{max}}$ oscillates and
then increases until the collapse of the HMNS sets in.  After the
collapse of the HMNS, $\rho_{\rm{max}}$ increases monotonically.  For
APR4-27, $\rho_{\rm{max}}$ also oscillates and increases soon after
the formation of the HMNS.  However it eventually reaches a relaxed
value, implying that a nearly stationary HMNS is the outcome.

Figure~\ref{type} summarizes the type of the merger process for each
EOS and for each mass.  We find that a HMNS is likely to be formed for
the merger with the EOS which provides a small compactness, $M/R$, for
each neutron star.  The reason is as follows.  In the case that the
compactness of each neutron star is small, two neutron stars merge at
a relatively large orbital separation.  As a result, the merged
neutron stars have large angular momentum at the onset of the merger,
which helps escaping the prompt collapse to a black hole.  The
long-lived HMNS is also likely to be formed for a total mass
$m_{0}\agt 2.7M_{\odot}$ with the EOS which has the maximum mass
exceeds $2M_{\odot}$, such as APR4, H4, and ALF2.

Note that it is practically impossible to precisely determine the
lifetime of the HMNS by the numerical simulation, because the HMNS
just before the collapse is marginally stable and its dynamics
depends strongly on a small perturbation and thus on the grid
resolution.  The lifetime also depends on the treatment of the thermal
effects~\cite{bib02}, which are determined by the value of
$\Gamma_{\rm{th}}$ in our simulations.  However we find that our
classification of the merger process depends very weakly on the grid
resolution and the value of $\Gamma_{\rm{th}}$.

\begin{table*}[htbp]
\begin{center}
\caption{The maximum mass, $M_{\rm{max}}$, the critical mass, 
$M_{\rm{crit}}$, and their ratio $k$ for each EOS.}\label{max}
 \begin{tabular}{l|cccccc} \hline \hline & APR4 & SLy & ALF2 & H3 & H4
 & PS \\ \hline $k$ & $1.3$ & $1.3$ & $1.4$ & $1.6$ & $1.5$ & $1.7$ \\
 $M_{\rm{max}}/M_{\odot}$ & 2.213 & 2.049 & 2.086 & 1.788 & 2.032 &
 1.755 \\ $M_{\rm{crit}}/M_{\odot}$ & $\simeq 2.8$ & $\simeq 2.8$ &
 $\simeq 3.0$ & $\simeq 2.9$ & $\gtrsim 3.0$ & $\simeq 3.0$\\ \hline
 \hline
\end{tabular}
\end{center}
\end{table*}

\subsubsection{\textit{The Critical mass}}

We define a critical mass, $M_{\rm{crit}}$, of binary neutron stars
for each EOS as follows: If the total mass of a binary neutron star,
$m_{0}$, exceeds $M_{\rm{crit}}$, a black hole is formed promptly
after the onset of the merger.

We write the value of $M_{\rm{crit}}$ in terms of $M_{\rm{max}}$ as
\begin{eqnarray}
M_{\rm{crit}}=k M_{\rm{max}},
\end{eqnarray}
where $k$ is a constant which depends on the EOS.  Table~\ref{max}
shows the maximum mass, the critical mass, and their ratio, $k$, for
each EOS together.  Here we find that $k$ is in the range,
\begin{eqnarray}
1.3\lesssim k \lesssim 1.7 ,
\end{eqnarray}
where $k\simeq 1.3$ for APR4 ($R_{1.4}=11.428$ km) and SLy
($R_{1.4}=11.736$ km) and $k\simeq 1.7$ for PS ($R_{1.4}=15.472$ km).
We find the correlation between $k$ and $R_{1.4}$, in which $k$ is
approximately an increasing function of $R_{1.4}$.  Note that the
results for APR4 and SLy agree with those in the previous
papers~\cite{bib31,bib29}.

Recent observation suggests $M_{\rm{max}}\geq 1.97\pm 0.04M_{\odot}$
in one $\sigma$ level~\cite{bib34}.  This fact and our results suggest
that a HMNS is likely to be formed transiently soon after the onset of
the merger for a binary neutron star merger with the typical total
mass, $m_{0}\simeq 2.7M_{\odot}$~\cite{bib41}.

\subsubsection{\textit{Final states: Black hole and Torus}}

\begin{figure}[htbp]
 \begin{center} \includegraphics[scale =0.7]{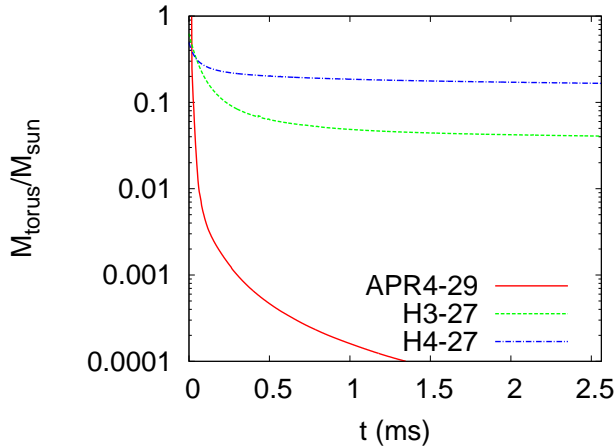}
 \end{center} \caption{The evolution of the torus mass,
 $M_{\rm{torus}}$, for three models.  The solid, dashed, and
 dash-dotted curves denote the results for models APR4-29 (type I),
 H3-27 (type II), and H4-27 (type III), respectively.  The time at the
 black hole formation is set to be $t=0$.  } \label{torusmass}
\end{figure}

As summarized in Sec. III.A.1, after the merger of a binary neutron
star, a black hole surrounded by a torus is eventually formed (see
Fig.~\ref{snapshot}).  Our interest here is to study how the amount of
the torus mass depends on the type of the merger process.
Figure~\ref{torusmass} shows the evolution of the torus mass for
models (types) APR4-29 (type I), H3-27 (type II), and H4-27 (type
III).  Here the time at the black hole formation is set to be $t=0$.
Table~\ref{t5.1} shows the results for the torus mass as well as the
type of the merger process for each model.  We find that the torus
mass is correlated to the type of the merger process as follows,
\begin{eqnarray}
M_{\rm{torus}}\lesssim 0.01M_{\odot} &   & \text{for type I},
\end{eqnarray}
\begin{eqnarray}
0.02M_{\odot} \lesssim M_{\rm{torus}} \lesssim 0.05M_{\odot} 
& \text{for type II},
\end{eqnarray}
\begin{eqnarray}
0.04M_{\odot}\lesssim M_{\rm{torus}}\lesssim 0.18M_{\odot} 
& \text{for type III}.
\end{eqnarray}
Here the torus mass is evaluated at about 1~ms after the black hole
formation.  Thus, the torus mass is larger in the case that a HMNS is
formed than that a black hole is formed promptly.  When a long-lived
HMNS is formed, in particular, the resulting torus mass could be
$M_{\rm{torus}}\geq 0.1M_{\odot}$ for many EOSs.  This feature can be
understood as follows.  In the HMNS, its envelope spins up because
angular momentum is transported from inner to outer parts of the HMNS
by gravitational torques caused by the non-axisymmetric structure.  As
a result, a part of the matter in the envelope of the HMNS, which has
sufficient angular momentum, does not fall into the black hole at the
collapse of the HMNS and it constitutes the torus.  However, we note
that the efficiency of the angular momentum transport depends on the
density profile and the degree of non-axisymmetry of the HMNS, which
depend on the EOS and the total mass.  For PS-27, indeed, the
resulting torus mass is only $\approx 0.04M_{\odot}$ which is much
less than those for APR4-27, H4-27, and ALF2-27.  Thus, for a special
EOS which leads to a large radius and a small maximum mass such as
$R_{1.4}\simeq 15.5$~km and $M_{\rm{max}}\simeq 1.76M_{\odot}$, the
torus mass may be small even if the merger process is type III.

Note that, in this study, we performed numerical simulations only for
the equal-mass systems.  For unequal-mass systems, a massive torus may
be formed even for the case that the merged neutron stars collapse
promptly to a black hole.  This is because the heavier star can
disrupt the less massive companion by tidal forces, in particular, for
high mass ratio and subsequent angular momentum transport in the
spiral arms formed from the tidally disrupted neutron star enhances
the torus formation, as indicated in Refs.~\cite{bib28, bib24, bib27}.

\begin{figure*}[htbp]
\begin{center}
\centerline{
\begin{tabular}{l l}
\rotatebox{0}{\includegraphics[scale=0.55]{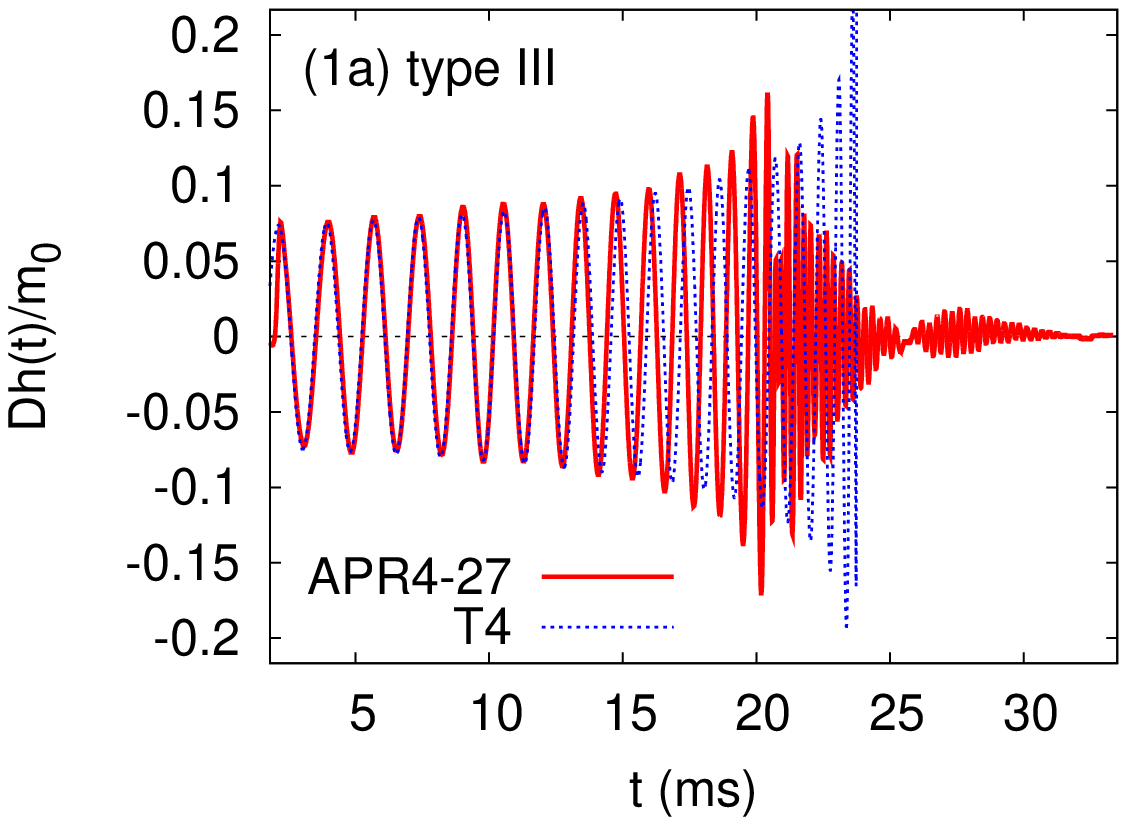}}
\rotatebox{0}{\includegraphics[scale=0.55]{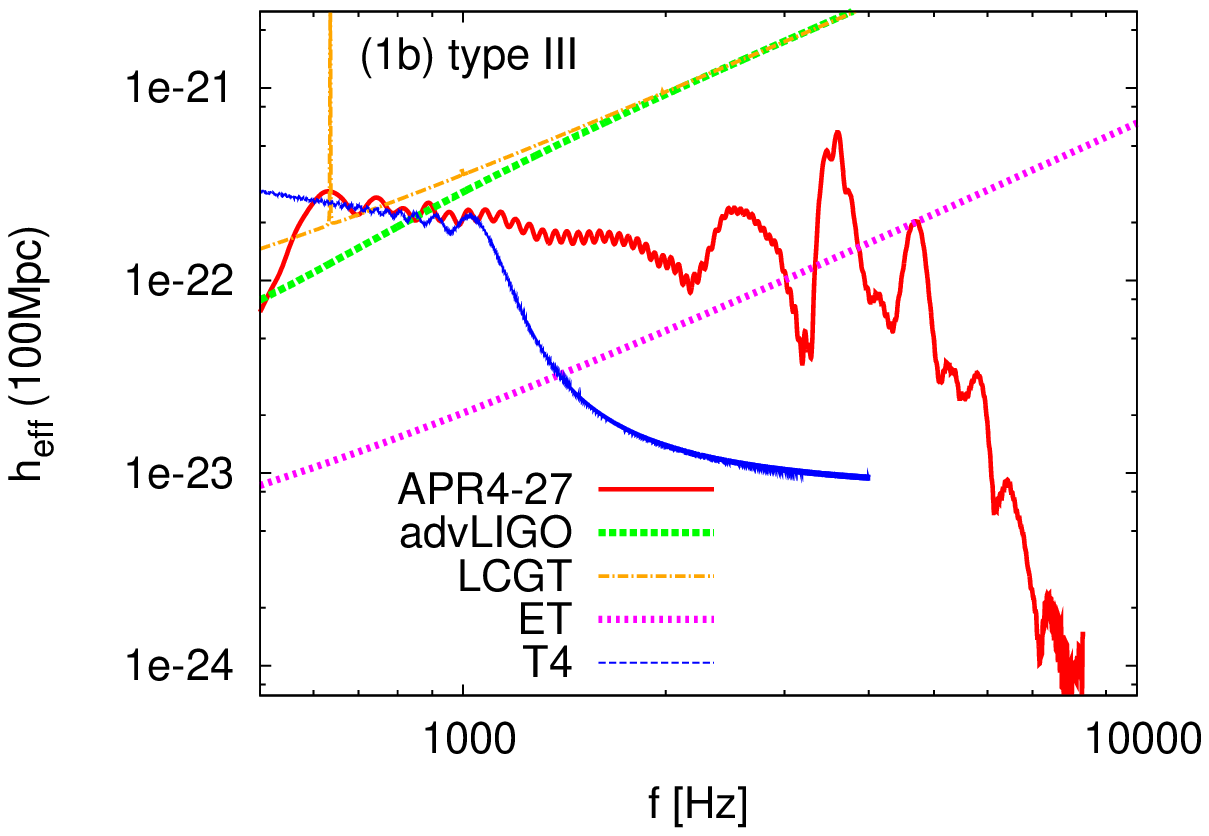}}\\
\rotatebox{0}{\includegraphics[scale=0.55]{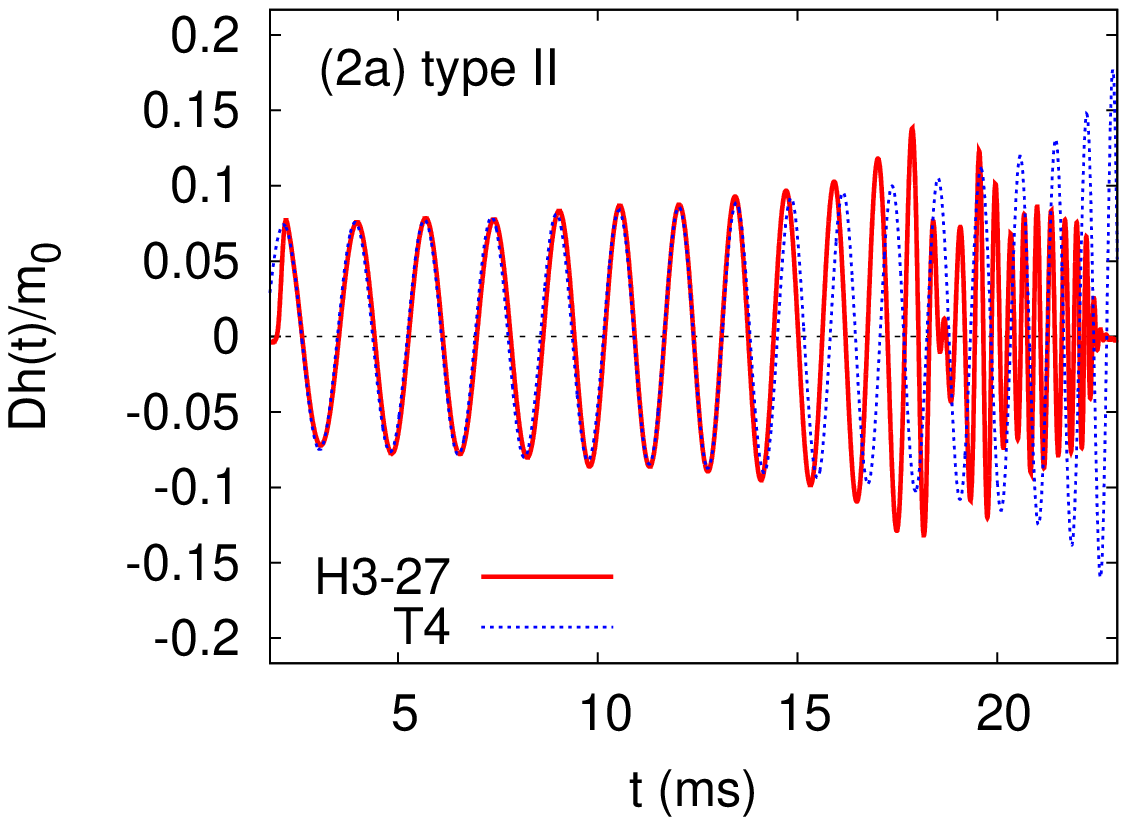}}
\rotatebox{0}{\includegraphics[scale=0.55]{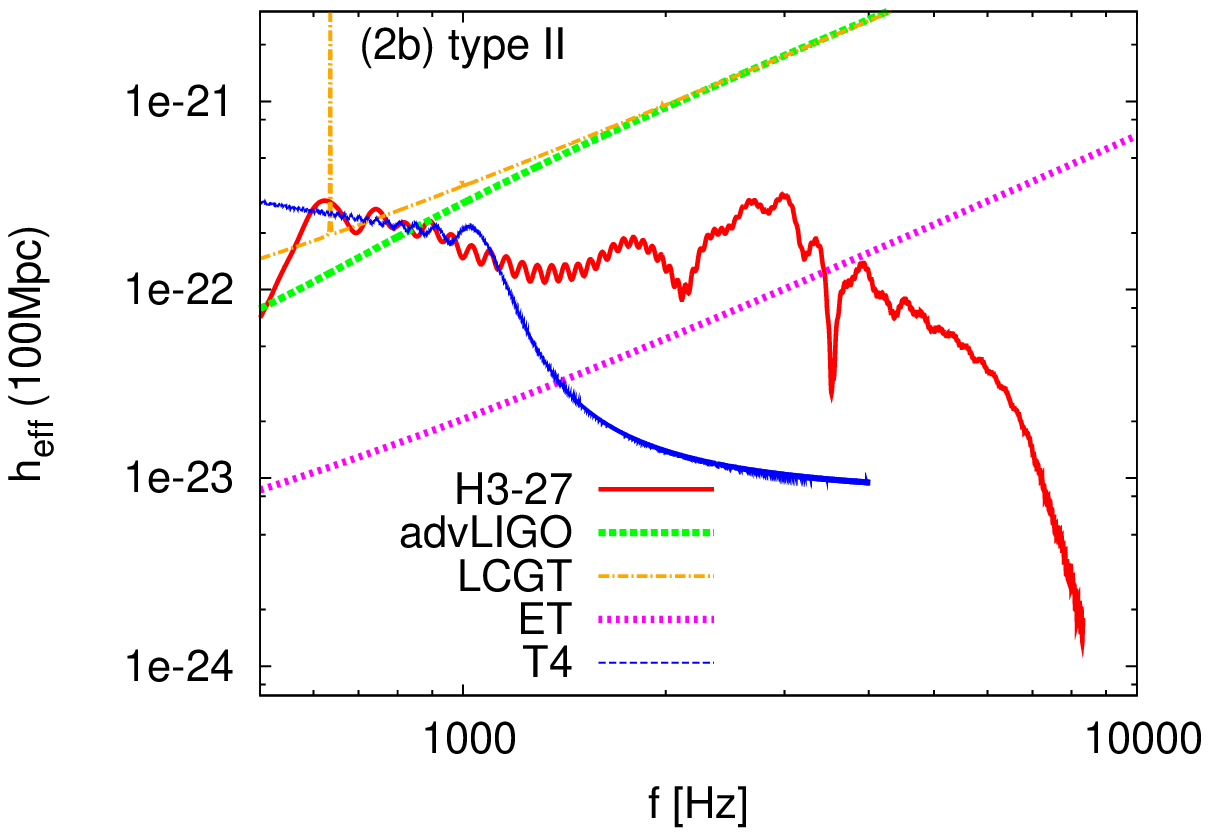}}\\
\rotatebox{0}{\includegraphics[scale=0.55]{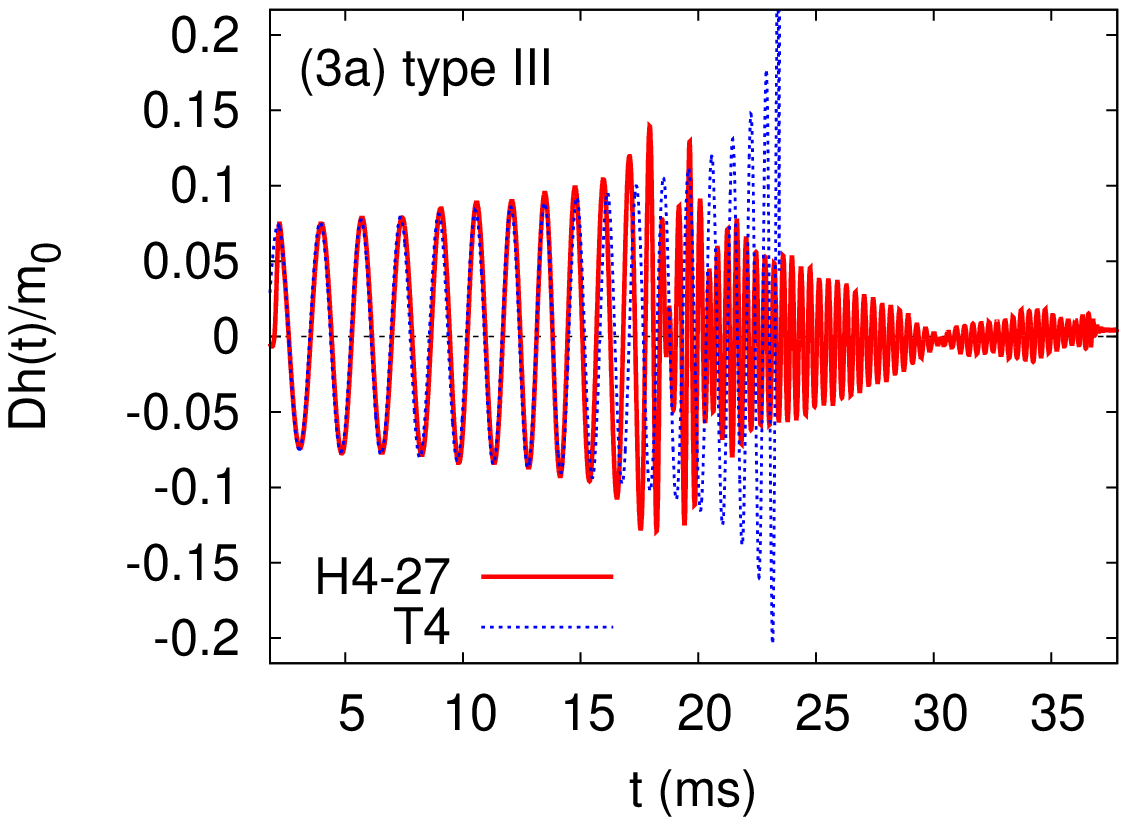}}
\rotatebox{0}{\includegraphics[scale=0.55]{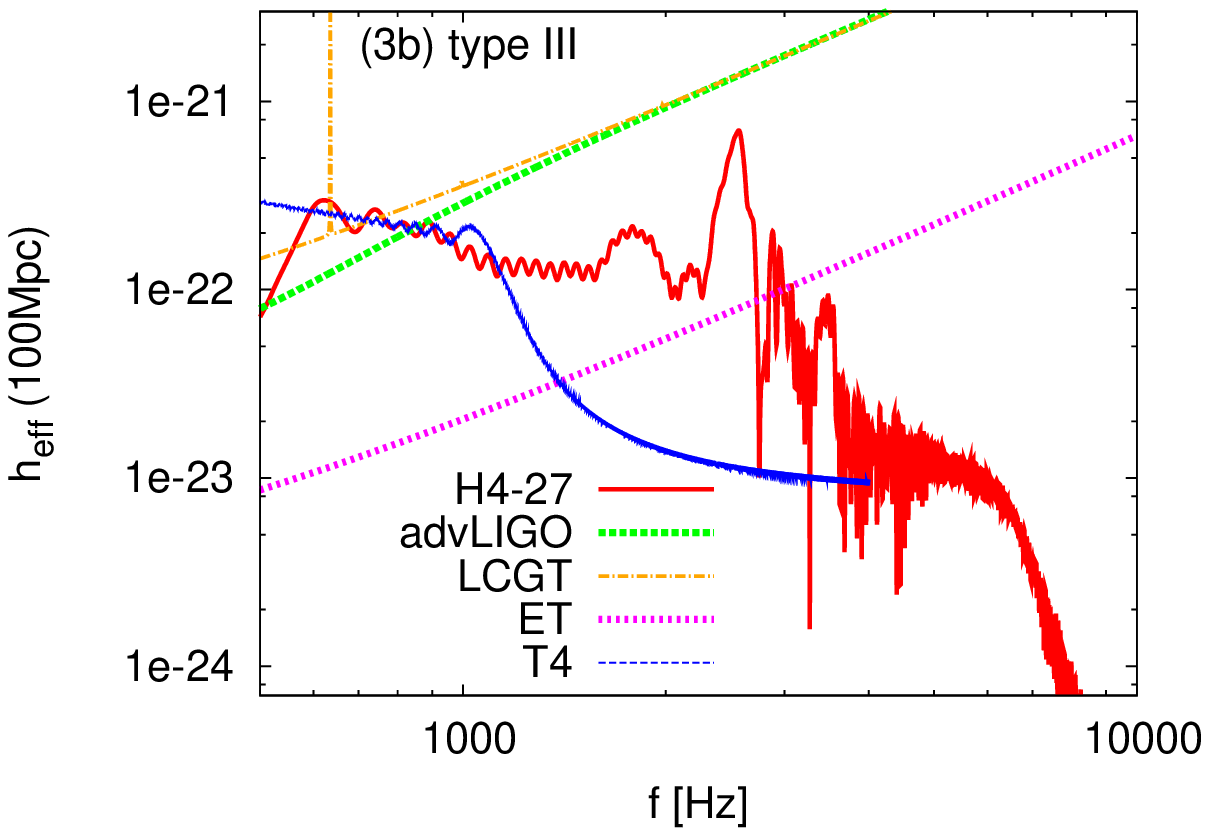}}\\
\rotatebox{0}{\includegraphics[scale=0.55]{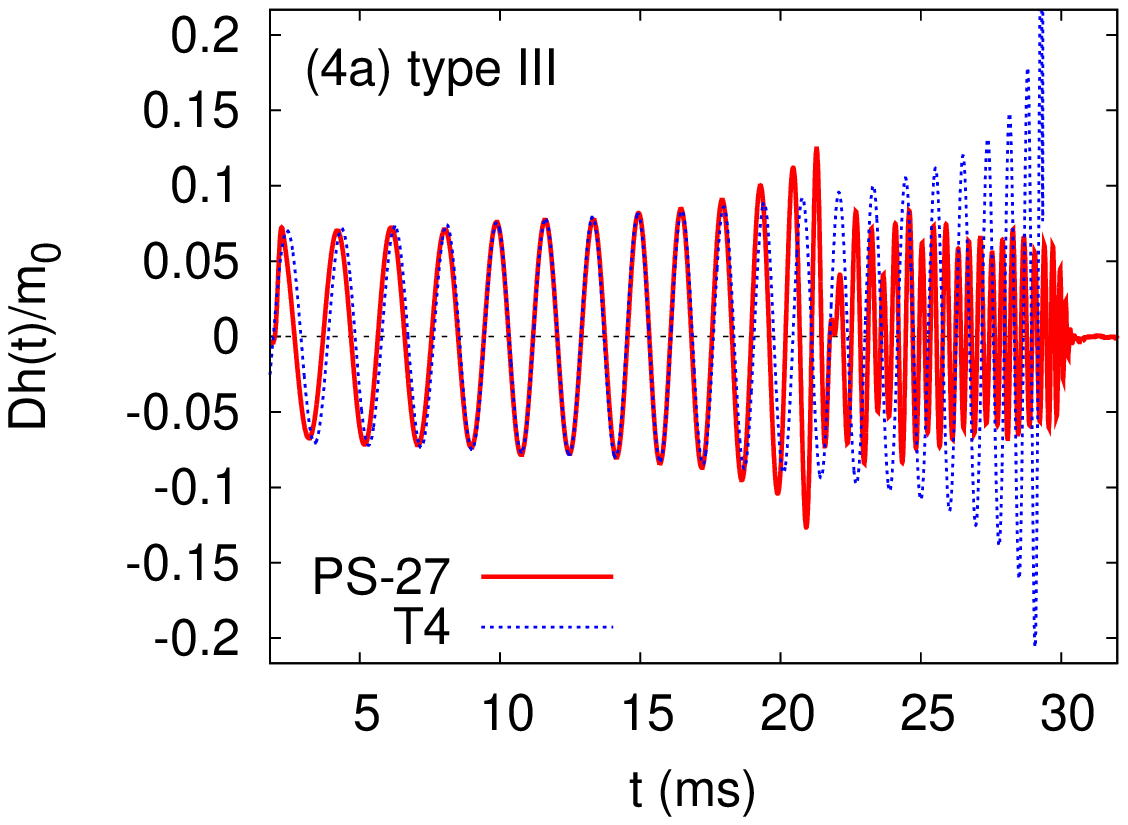}}
\rotatebox{0}{\includegraphics[scale=0.55]{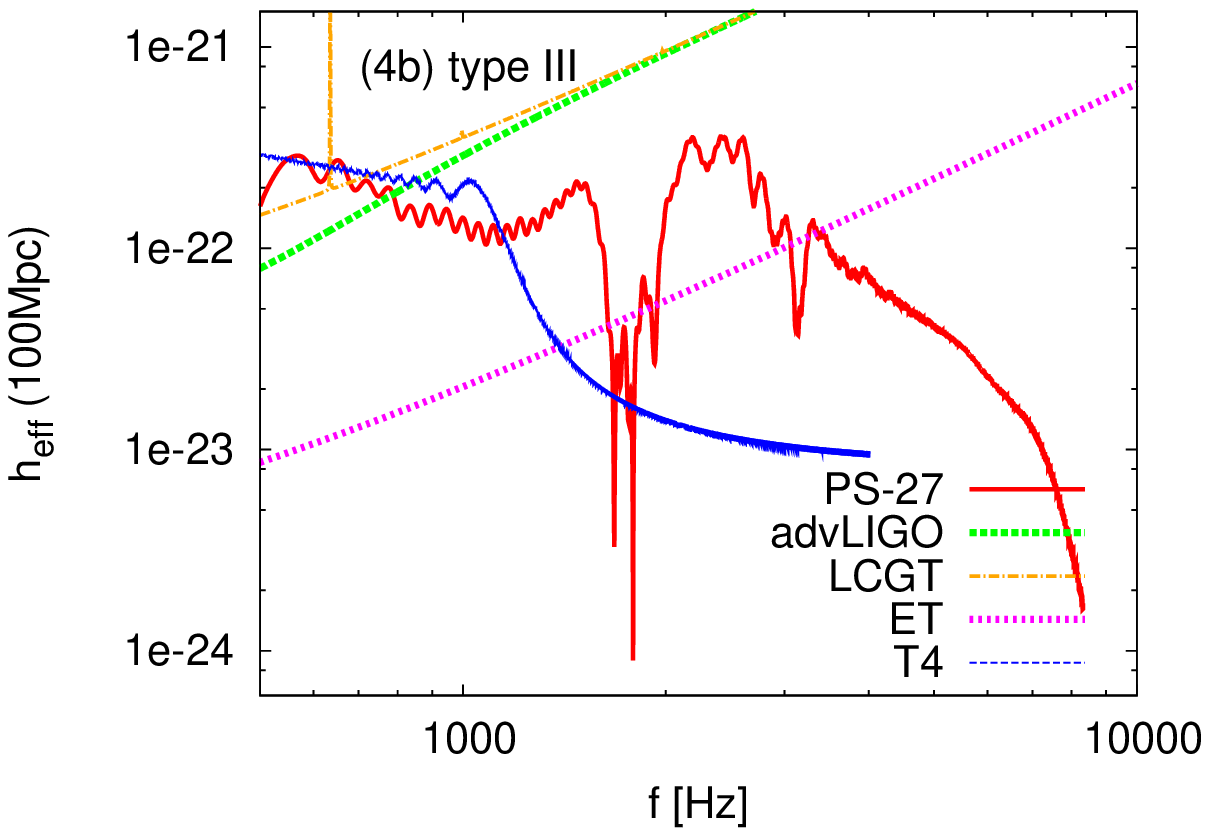}}
\end{tabular}
}
\caption{
Gravitational waveforms and their spectra. The solid and dashed curves
in the left panels denote the waveforms calculated by the simulation
and Taylor T4 formula, respectively.  The solid and dashed curves in
the right panels denote the spectra calculated by the simulation, and
spectrum calculated by Taylor T4 formula, respectively, at a
hypothetical source distance of 100~Mpc.  The effective amplitude for
the most optimistic direction of the source is shown.  Here the noise
levels of advanced LIGO (Optimal NSNS version), LCGT (Broadband
version), and Einstein Telescope are shown together.  (1a) and (1b)
for APR4-27 (type III), (2a) and (2b) for H3-27 (type II), (3a) and
(3b) for H4-27 (type III), (4a) and (4b) for PS-27 (type III). }
\label{gw}
\end{center}
\end{figure*}

\begin{figure*}[htbp]
\begin{center}
\centerline{
\begin{tabular}{l l}
\rotatebox{0}{\includegraphics[scale=0.55]{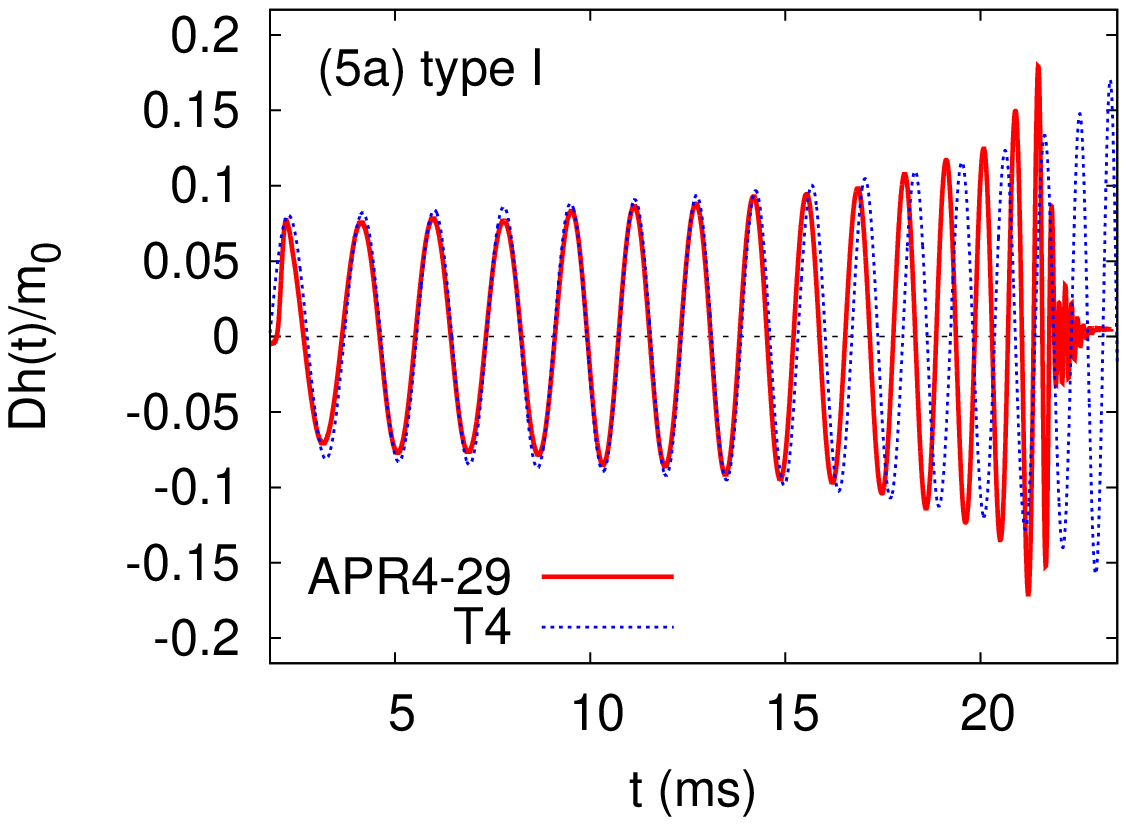}}
\rotatebox{0}{\includegraphics[scale=0.55]{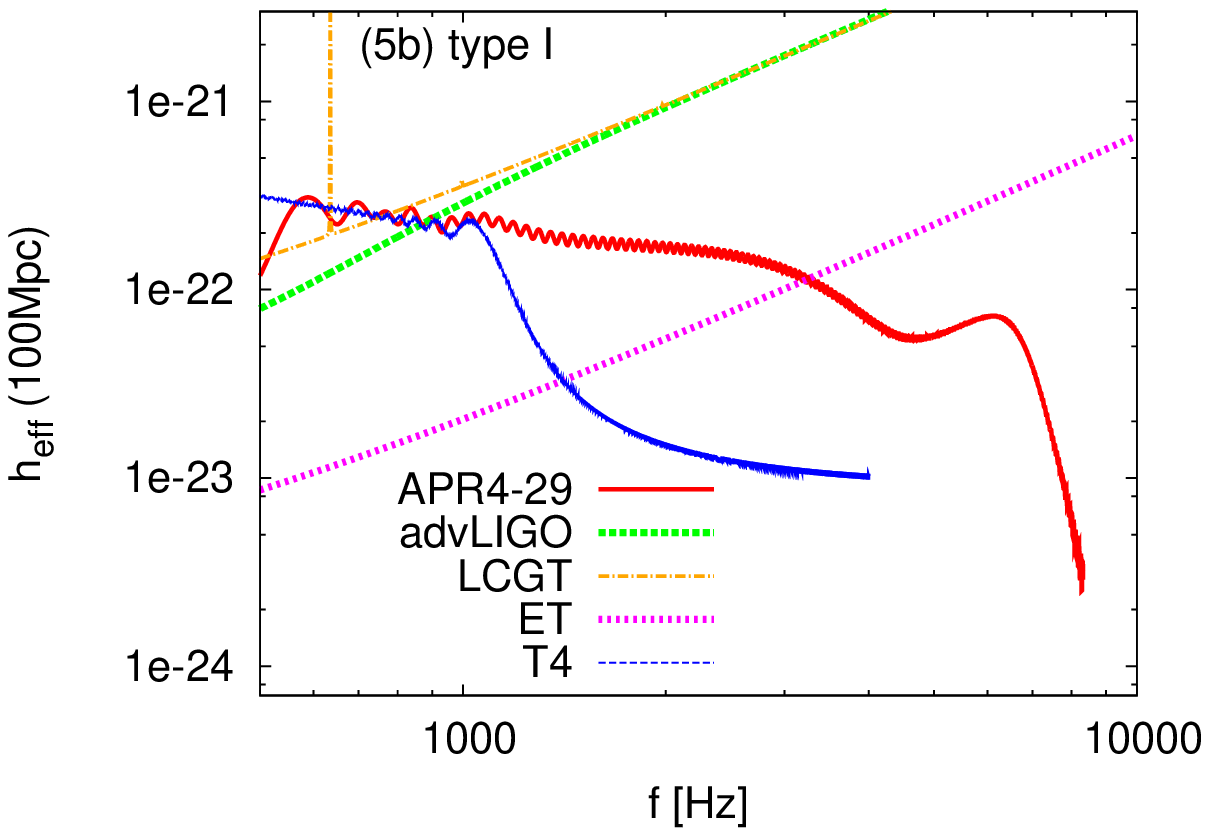}}\\
\rotatebox{0}{\includegraphics[scale=0.55]{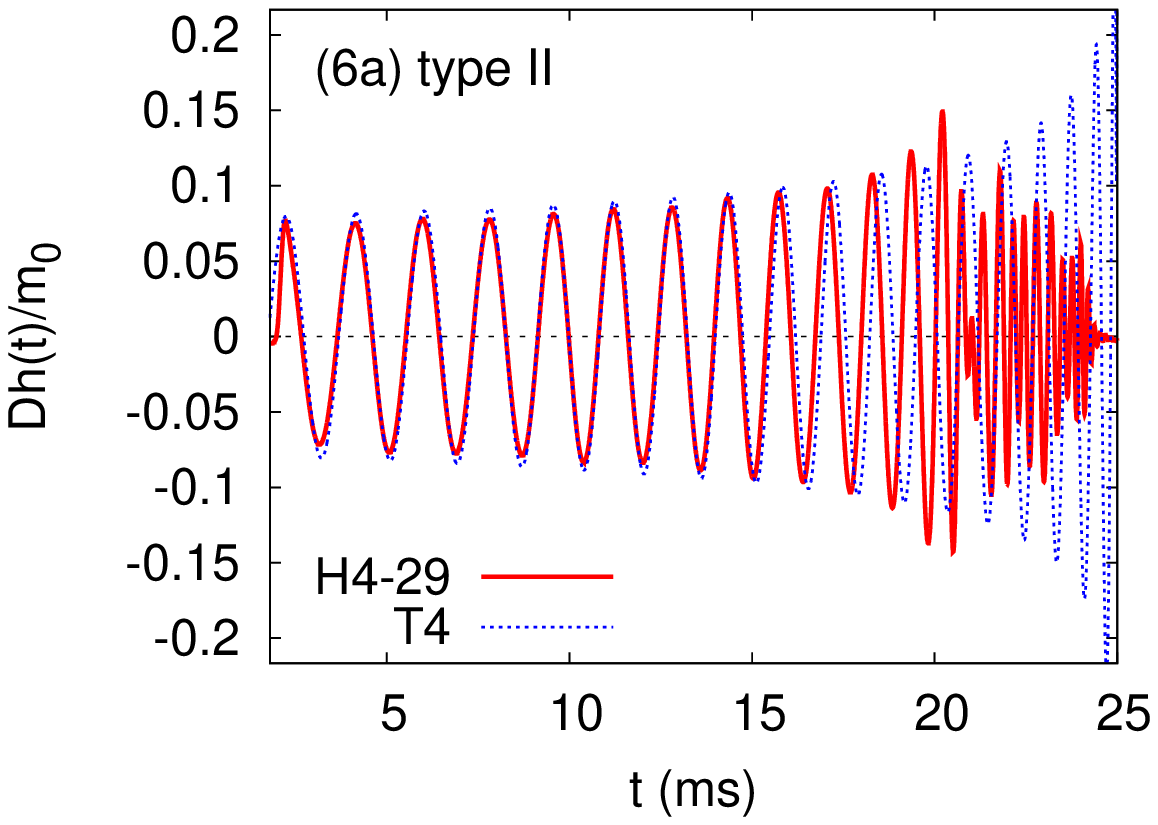}}
\rotatebox{0}{\includegraphics[scale=0.55]{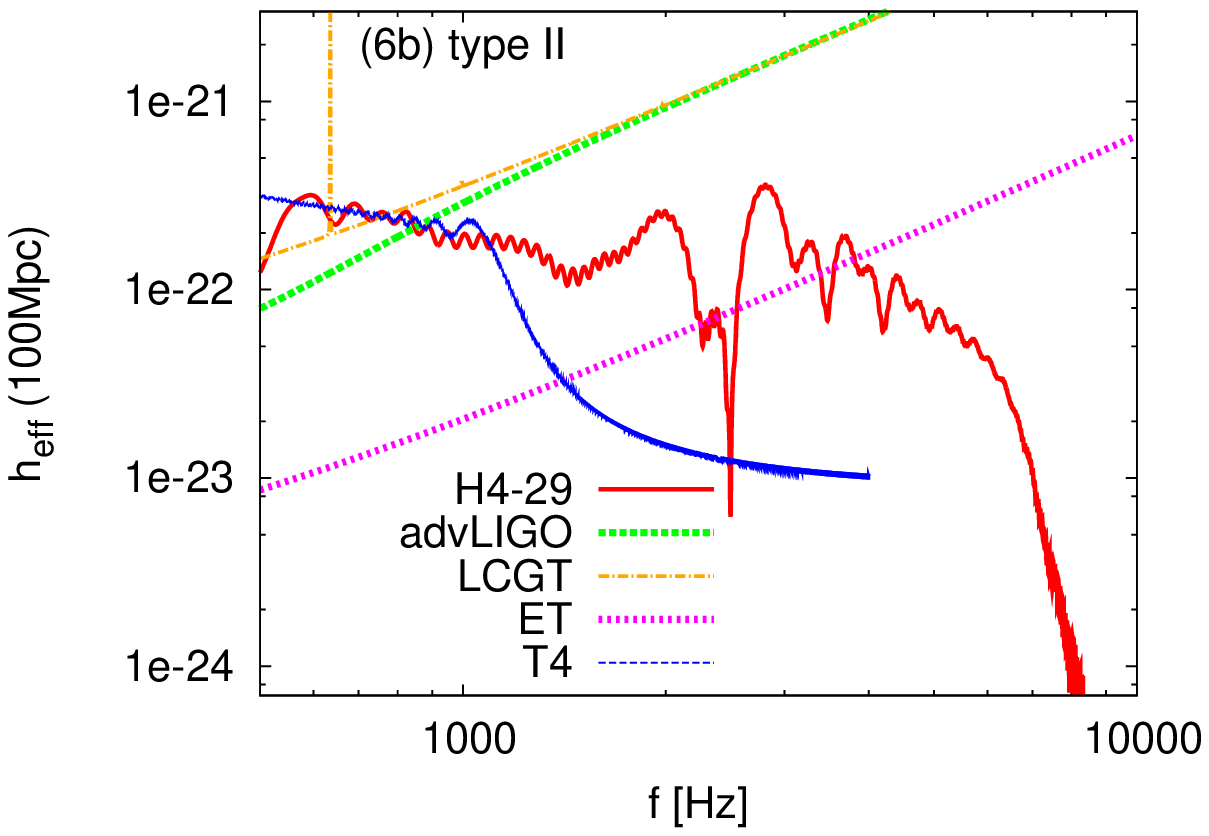}}
\end{tabular}
}
\caption{The same as Fig.~\ref{gw} but for $m_{0}=2.9M_{\odot}$.
(5a) and (5b) for APR4-29 (type I),
(6a) and (6b) for H4-29 (type II). }
\label{gw2}
\end{center}
\end{figure*}

\subsection{Gravitational Waves}

Gravitational waves are emitted during the merger until a stationary
black hole is formed.  The gravitational waveforms reflect the
dynamical behavior of the merger process.  In the following, we
classify the gravitational waveforms and their spectra into three
types in the same way as the merger process, and discuss their
features.

\subsubsection{\textit{Gravitational waveforms}}

The gravitational waveforms for models APR4-27, H3-27, H4-27, and
PS-27 together with post-Newtonian waveforms, calculated by the
so-called Taylor T4 formula~\cite{bib26}, are shown in Figs.~\ref{gw}
(1a) -- (4a).  For more massive models APR4-29 and H4-29 the
gravitational waveforms are shown in Figs.~\ref{gw2} (5a) and (6a).
The gravitational waveform for the inspiral phase is similar for all
the models.  In the late inspiral stage, the gravitational-wave phase
starts deviating from the post-Newtonian one at $3$ -- $5$~ms before
the onset of the merger.  This is because the effects of the finite
size of two neutron stars play an important role for their orbital
motion.  A detailed analysis for the finite-size effect will be
published in a future paper.  In the following, we focus on the
gravitational waveform of the merger and the ringdown phases.

\textit{Type I}. The gravitational waveform for APR4-29 is shown 
in Fig.~\ref{gw2} (5a).  The amplitude of gravitational waves
increases gradually in the inspiral phase until the merger sets in at
$t\simeq 22$~ms.  Soon after the onset of the merger, ringdown
gravitational waves are emitted by the oscillating black hole for
$\sim 1$~ms.  Then the amplitude approaches zero because the black
hole becomes stationary.

\textit{Type II}.
A short-lived HMNS is formed after the onset of the merger.  Then the
oscillating and rotating HMNS emits quasiperiodic gravitational waves.
The gravitational waveform of H3-27 is shown in Fig.~\ref{gw} (2a).
In this case, two neutron stars merge at $t\simeq 18$ ms, at which the
amplitude of gravitational waves is small transiently.  This implies
that the merged object has a nearly axisymmetric ellipsoidal shape at
the moment.  After this, the core bounces and the HMNS is formed,
which has a double-core structure.  Then quasiperiodic gravitational
waves with a high amplitude are emitted.  At $t\simeq 23$ ms, the
amplitude of gravitational waves damps suddenly.  This is because the
HMNS collapses to a black hole before the HMNS becomes an axisymmetric
ellipsoidal shape.  A similar waveform is also seen for H4-29 in
Fig.~\ref{gw2} (6b).

\textit{Type III}. 
The gravitational waveform of H4-27 is shown in Fig.~\ref{gw} (3a).
In this case, two neutron stars merge at $t\simeq 18$ ms, at which the
amplitude of gravitational waves is small transiently as in H3-27.
After this, a non-axisymmetric HMNS of the double-core structure is
formed and it emits quasiperiodic gravitational waves.  For $t \agt
30$~ms, gravitational waves of small amplitude is emitted
quasi-stationarily until $t \sim 37$ ms.  During this phase, the HMNS
has a nearly axisymmetric ellipsoidal shape.  At $t\simeq 37$ ms, the
HMNS collapses to a black hole and the amplitude of gravitational
waves damps eventually.

For APR4-27 (see Fig.~\ref{gw} (1a)), an ellipsoidal HMNS is formed
after the onset of the merger, and thus, quasiperiodic gravitational
waves are emitted for the first $\sim 10$ ms after the formation of
the HMNS.  The HMNS loses angular momentum due to the gravitational
radiation reaction and the ellipticity gradually decreases, resulting
in the decrease of the gravitational-wave amplitude.  Thus, the
gravitational waveform is similar to that for H4-27.  However, the
lifetime of the HMNS for APR4-27 is much longer than 15 ms.  Thus we
did not follow the collapse of the HMNS for APR4-27.

For PS-27 (see Fig.~\ref{gw} (4a)), the gravitational waveform is
different from those of H4-27 and APR4-27.  In this case, it is
similar to those for H3-27 and H4-29: A HMNS of double-core structure
is formed after the onset of the merger, and emits quasiperiodic
gravitational waves.  After a substantial emission of gravitational
waves, the HMNS collapses to a black hole before it becomes a nearly
axisymmetric ellipsoid.
        
\subsubsection{\textit{Spectra}}

The right panels of Figs.~\ref{gw} and \ref{gw2} show the spectra of
gravitational waves for models APR4-27, H3-27, H4-27, PS-27, APR4-29,
and H4-29 together with the spectrum calculated by the post-Newtonian
approximation (Taylor T4), and the sensitivity curves of advanced LIGO
(Optimal NSNS version)~\cite{bib52}, of LCGT (Broadband
version)~\cite{bib08}, and of Einstein telescope~\cite{bib53}.  Here
we assume that gravitational waves are observed at a distance of
$100$~Mpc from the sources along the most optimistic source direction.
The common feature for each type is that the effective amplitude,
$h_{\rm{eff}}$, decreases with the increase of $f$ until $f \simeq
1$~kHz, in the inspiral phase.  The spectrum shape above $f\simeq
7$~kHz is also qualitatively the same irrespective of the model, which
is caused by the quasi-normal-mode oscillations of the remnant black
hole.  However, the spectrum shape between $1$~kHz and $7$~kHz, for
which gravitational waves are emitted in the merger phase, depends
strongly on the EOS and on the total mass as summarized in the
following:

\textit{Type I}. Because no HMNS is formed, the shape of the spectrum 
is quite simple (see Fig.~\ref{gw2} (5b)).  The effective amplitude
decreases monotonically until $f\simeq 5$~kHz.  Note that the power of
the spectrum shape changes at $f\simeq 3$~kHz.  This frequency is
called the cut-off frequency, which is related to the compactness of
two neutron stars (see Ref.~\cite{bib24} for details).  The bump
around $f \simeq 6$~kHz is caused by the rotation of the merged object
just before the collapse to a black hole (see Fig.~\ref{snapshot},
\textit{top center}).  We find that this bump is enhanced in the case
that the total mass is close to the critical mass, $m_{0}\simeq
M_{\rm{crit}}$. 

\textit{Type II}. 
A spectrum shape of type II is shown in Figs.~\ref{gw} (2b) and
\ref{gw2} (6b).  In this case, we find several peaks caused by the
rotation and oscillation modes of the HMNS.  There is the maximum peak
at $f \simeq 2.5$--$3$~kHz caused by the rotation of the HMNS.  The
frequency of this peak is higher for the more compact HMNS, because
the angular velocity of the merged neutron stars is approximately
written by $\Omega \propto m_{0}^{1/2}/R_{\rm{ns}}^{3/2}$ at the onset
of the merger. Here $R_{\rm{ns}}$ is the radius of two neutron stars.
However, the peaks in the spectrum are too small to be detected by the
advanced detectors such as advanced LIGO and LCGT for $D=100$~Mpc,
because the lifetime of the HMNS is short and the accumulated
gravitational-waves cycles are small.  There is a shallow decay of the
spectrum around $4
\lesssim f \lesssim 7$~kHz.  This is the common feature in the case
that a HMNS is formed (see Figs.~\ref{gw} (2b)--(4b), and \ref{gw2}
(6b)).

\textit{Type III}. Figures~\ref{gw} (1b) and \ref{gw} (3b) show the 
spectrum shape of APR4-27 and H4-27.  We find that the peak amplitude
of the spectrum at $f\simeq 2.5$~kHz is larger than that for type II
and the primary oscillation mode appears clearly.  Note that the
amplitude of the peak is slightly smaller than the sensitivity curve
of advanced LIGO (Optimal NSNS version) and LCGT (Broadband version)
for a hypothetical distance of 100~Mpc.  If this peak could be
observed with optimized detectors or for an event of short distance,
we would get information about the physics of HMNSs and could
constrain the supernuclear-density EOS~\cite{bib25}.

Note that the spectrum shape for type III depends on the EOS.  For
example, the spectrum shape of PS-27 is different from that for
APR4-27 and H4-27 (see Fig.~\ref{gw} (1b), \ref{gw} (3b), and
Fig.~\ref{gw} (4b)).  The bump around the maximum peak at $f\simeq
2.5$~kHz in the spectrum of PS-27 is broader than that of H4-27.

\section{Summary and Discussion}

We studied the dependence of the dynamical behavior of the binary
neutron star merger on the EOS of the supernuclear-density matter in
numerical relativity with piecewise-polytropic EOSs.  We performed
numerical simulations for 6 stiff EOSs and for several total masses in
the range of $2.7M_{\odot}$--$3.0M_{\odot}$.  For all the cases, the
equal-mass binary system was considered.

We classified the merger process into three types: (i) a black hole is
promptly formed (type I); (ii) a short-lived HMNS is formed (type II);
(iii) a long-lived HMNS is formed (type III).  For a given total mass
of the binary neutron star, the type of the merger process depends
strongly on the EOS.  In particular, the compactness of the neutron
stars and the maximum mass of spherical neutron stars for a given EOS
are key quantities for determining whether a HMNS is formed
transiently or not.  We found that the critical mass of the prompt
formation of a black hole is in the range of $1.3 \lesssim
M_{\rm{crit}}/M_{\rm{max}} \lesssim 1.7$.  The latest observation of
the high-mass neutron star with mass $1.97\pm 0.04 M_{\odot}$ suggests
that $M_{\rm{crit}}$ is larger than $2.7M_{\odot}$.  These facts
indicate that a HMNS should be formed at least transiently for the
total mass of the binary neutron stars, $m_{0}\sim 2.6M_{\odot}$ or
less.  We found that a long-lived HMNS is formed for $m_{0}\sim
2.7M_{\odot}$ and for the EOS with which the maximum mass of spherical
neutron stars exceeds $2M_{\odot}$, such as APR4 and H4.

We studied the properties of a torus around a black hole formed after
the merger.  We found that the torus mass depends strongly on the type
of the merger process.  Specifically, the range of the torus mass is
$M_{\rm{torus}} \lesssim 0.01M_{\odot}$ for type I, $0.02M_{\odot}
\lesssim M_{\rm{torus}}\lesssim 0.05M_{\odot}$ for type II, and $0.04
\lesssim M_{\rm{torus}}\lesssim 0.18M_{\odot}$ for type III.  Thus we
found that the torus mass is larger in the case that the HMNS is
formed than in the case that a black hole is formed promptly.  This is
because materials in the outer envelope of the HMNS receive angular
momentum from the central part via gravitational torques which work on
the non-axisymmetric ellipsoidal HMNS.  As a result, a part of the
materials in the outer envelope does not fall into the black hole soon
after the collapse of the HMNS and remains around the black hole to be
a torus.  In this sense, we conclude that the HMNS will play an
important role for the merger scenario of short GRBs.  Note that we
performed numerical simulations only for the equal-mass system.  For
the unequal-mass case, the dynamical evolution of the merger could
be modified.  The heavier star may tidally disrupt the less massive
companion if the mass ratio is high.  As a result, the torus mass
around the black hole may be enhanced (see Refs.~\cite{bib28, bib24,
bib27}).

We also studied gravitational waves from the binary neutron star
merger.  The gravitational waveforms and their spectra depend strongly
on the merger process.  For type II and type III, we found that a
large amount of gravitational waves are emitted by the HMNS which has
a non-axisymmetric configuration.  With decreasing its angular
momentum, the HMNS approaches an axisymmetric ellipsoid and the
amplitude of gravitational waves decreases.

The amplitude of the spectrum for type I decreases monotonically with
increasing $f$ until $3$--$4$~kHz because the HMNS is not formed.
Note that there is a bump at $f \sim 5$--$6$~kHz caused by
gravitational waves emitted by a merged object just before the
collapse.  By contrast, for type II and type III, the spectra have a
complicated shape caused by the oscillation and rotation of the HMNS.
At $f\sim 2$--$3$~kHz, there is the maximum peak cause by the
fundamental rotation mode of the non-axisymmetric HMNS.  We also found
that several peaks due to the oscillation mode coupling with the rotation
exist at both sides of the maximum peak.

The detectability of gravitational waves from the HMNS is estimated.
With advanced detectors such as advanced LIGO and LCGT, we may be able
to detect gravitational waves caused by the rotation mode of the very
long-lived HMNS ($\tau_{H}\geq 10$~ms) at around $f\sim 2$--$3$~kHz if
the distance to the source is much smaller than 100~Mpc.  If these
waves are observed, we will get information about the physics of HMNSs
and may be able to constrain properties of the supernuclear-density
matter~\cite{bib25}. 

\begin{acknowledgments}
We thank Y. Sekiguchi, Y. Suwa, and T. Nakamura for useful discussions
and comments.  This work was supported by Grant-in-Aid for Scientific
Research (21340051), by Grant-in-Aid for Scientific Research on
Innovative Area (20105004), by the Grant-in-Aid of JSPS, by HPCI
Strategic Program of Japanese MEXT, and by Grant-in-Aid for Young
Scientists (B) 22740178.
\end{acknowledgments}

\end{document}